\documentclass[journal]{IEEEtran}
\usepackage{cite}
\usepackage{amsmath,amssymb,amsfonts}
\usepackage{algorithmic}
\usepackage{graphicx}
\usepackage{textcomp}
\usepackage{color}
\usepackage{subcaption}

\ifCLASSINFOpdf
\else
\fi

\begin{document}

\title{Residual Attention U-Net for Automated Multi-Class Segmentation of COVID-19 Chest CT Images}

\author{Xiaocong Chen, Lina Yao, \IEEEmembership{Member, IEEE}, and Yu Zhang, \IEEEmembership{Senior Member, IEEE}
\thanks{X. Chen, L.Yao are with the School of Computer Science and Engineering at University of New South Wales, NSW 2052, Australia (e-mail: xiaocong.chen@unsw.edu.au, lina.yao@unsw.edu.au). }
\thanks{Y. Zhang is with the Department of Psychiatry and Behavioral Sciences at Stanford University, CA 94305, USA (email: yzhangsu@stanford.edu).}}

%
%

\markboth{Journal of \LaTeX\ Class Files,~Vol.~14, No.~8, August~2015}%
{Shell \MakeLowercase{\textit{et al.}}: Bare Demo of IEEEtran.cls for IEEE Journals}
%



\maketitle

\begin{abstract}
The novel coronavirus disease 2019 (COVID-19) has been spreading rapidly around the world and caused significant impact on the public health and economy. However, there is still lack of studies on effectively quantifying the lung infection caused by COVID-19. As a basic but challenging task of the diagnostic framework, segmentation plays a crucial role in accurate quantification of COVID-19 infection measured by computed tomography (CT) images. To this end, we proposed a novel deep learning algorithm for automated segmentation of multiple COVID-19 infection regions. Specifically, we use the Aggregated Residual Transformations to learn a robust and expressive feature representation and apply the soft attention mechanism to improve the capability of the model to distinguish a variety of symptoms of the COVID-19. With a public CT image dataset, we validate the efficacy of the proposed algorithm in comparison with other competing methods. Experimental results demonstrate the outstanding performance of our algorithm for automated segmentation of COVID-19 Chest CT images. Our study provides a promising deep leaning-based segmentation tool to lay a foundation to quantitative diagnosis of COVID-19 lung infection in CT images.
\end{abstract}


\section{Introduction}
\label{sec:introduction}
The novel coronavirus disease 2019, also known as COVID-19 outbreak first noted in Wuhan in the end of 2019, has been spreading rapidly worldwide \cite{zhu2020novel}. As an infectious disease, COVID-19 is caused by severe acute respiratory syndrome coronavirus and presents with symptoms including fever, dry cough, shortness of breath, tiredness and so on. As the April 9th, over 1.5 million people around the world have been confirmed as COVID-19 infection with a case fatality rate of about 5.7 \% according to the statistic of World Health Organization\footnote{https://www.who.int/emergencies/diseases/novel-coronavirus-2019/situation-reports}.

So far, no specific treatment has proven effective for COVID-19. Therefore, accurate and rapid testing is extremely crucial for timely prevention of COVID-19 spread. Real-time reverse transcriptase polymerase chain reaction (RT-PCR) has been referred as the standard approach for testing COVID-19. However, RT-PCR testing is time-consuming and limited by the lack of supply test kits \cite{shan+2020lung,narin2020automatic}. Moreover, RT-PCR has been reported to suffer from low sensitivity and repeated checking is typically needed for accurate confirmation of a COVID-19 case. This indicates that many patients will not be confirmed timely \cite{long2020diagnosis,ai2020correlation}, thereby resulting in a high risk of infecting a larger population.

In recent years, imaging technology has emerged as a promising tool for automatic quantification and diagnosis of various diseases. As a routine diagnostic tool for pneumonia, chest computed tomography (CT) imaging has been strongly recommended in suspected COVID-19 cases for both initial evaluation and follow-up. Chest CT scans were found very useful in detecting typical radiographic features of COVID-19 \cite{li2020coronavirus}. A systematic review \cite{salehi2020coronavirus} concluded that CT imaging of chest was found sensitive for checking COVID-19 even before some clinical symptoms were observed. Specifically, the imaging features including ground class opacification, consolidation, and pleural effusion have been frequently observed in the chest CT images scanned from COVID-19 patients \cite{huang2020clinical,wang2020review,shi2020review}.

Accurate segmentation of these important radiographic features is crucial for a reliable quantification of COVID-19 infection in chest CT images. Segmentation of medical imaging needs to be manually annotated by well-trained expert radiologists. The rapidly increasing number of infected patients has caused tremendous burden for radiologists and slowed down the labeling of ground-truth mask. Thus, there is an urgent need for automated segmentation of infection regions, which is a basic but arduous task in the pipeline of computer-aided disease diagnosis \cite{gaal2020attention}. However, automatically delineating the infection regions from the chest CT scans is considerably challenging because of the large variation in both position and shape across different patients and low contrast of the infection regions in CT images \cite{shan+2020lung}. 

Machine learning-based artificial intelligence provides a powerful technique for the design of data-driven methods in medical imaging analysis \cite{shi2020review}. Developing advanced deep learning models would bring unique benefits to the rapid and automated segmentation of medical images \cite{shen2017deep}. So far, fully convolutional networks have proven superiority over other widely used registration-based approaches for segmentation \cite{gaal2020attention}. In particular, U-Net models work decently well for most segmentation tasks in medical images \cite{ronneberger2015u,baumgartner2017exploration,alom2018recurrent,shan+2020lung}. However, several potential limitations of U-Net have not been effectively addressed yet. For example, the U-Net model is hard to capture the complex features such as multi-class image segmentation and recover the complex feature into the segmentation image~\cite{oktay2018attention}.  
There are also a few successful applications that adopt U-Net or its variants to implement the CT image segmentation, including heart segmentation~\cite{ye2019multi}, liver segmentation~\cite{liu2019liver}, or multi-organ segmentation~\cite{dong2019automatic}.
However, segmentation of COVID-19 infection regions with deep learning remains under explored. The COVID-19 is a new disease but very similar with the common pneumonia in the medical imaging side, which makes its accurate quantification considerably challenging.
Recent advancement of the deep learning method provide heaps of insightful ideas about improving the U-Net architecture. The most popular one is the deep residual network (ResNet)~\cite{he2016deep}. ResNet provided an elegant way to stacked CNN layers and demonstrate the strength when combined with U-Net \cite{ibtehaz2020multiresunet}. On the other hand, attention was also applied to improve the U-Net and other deep learning models to boost the performance~\cite{oktay2018attention,chen2018interpretable}.

Accordingly, we propose a novel deep learning model for rapid and accurate segmentation of COVID-19 infection regions in chest CT scans. Our developed model is based on the U-Net architecture, inspired with recent advancement in the deep learning field. We exploit both the residual network and attention mechanism to improve the efficacy of the U-Net. Experimental analysis is conducted with a public CT image dataset collected from patients infected with COVID-19 to assess the efficacy of the developed model. The outstanding performance demonstrates that our study provides a promising segmentation tool for the timely and reliable quantification of lung infection, toward to developing an effective pipeline for precious COVID-19 diagnosis.

The rest of the paper is summarized as follows. We first review some related work about existing deep learning methods for CT image segmentation in Section II Related Work. Our proposed new deep learning model is detailedly described in Section III Methodology, including the U-Net structure, the methods used to improve the encoder and decoder. The experimental study and performance assessment are described in Section IV, followed by discussion and summary of our study.
\section{Methodology}
\begin{figure*}[ht]
    \centering
    \includegraphics[width=0.8\linewidth]{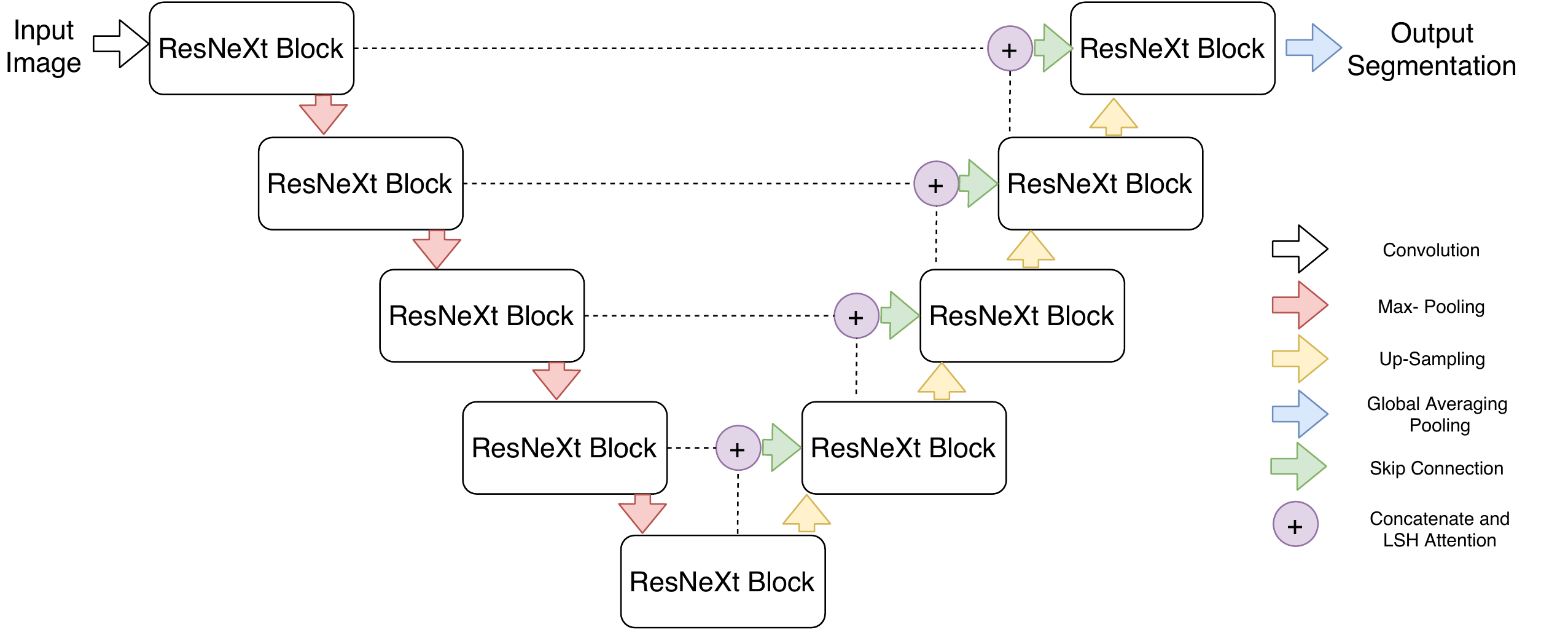}
    \caption{Illustration of our developed residual attention U-Net model. The aggregated ResNeXt blocks are used to capture the complex feature from the original images. The left side of the U-Shape serves as encoder and the right side as decoder. Each block of the decoder receives the feature representation learned from the encoder, and concatenates them with the output of deconvolutional layer followed by LSH attention mechanism. The filtered feature representation after the attention mechanism is propagated through the skip connection.}
    \label{fig:model}
\end{figure*}

This section will introduce our proposed Residual Attention U-Net for the lung CT image segmentation in detail. We start by describing the overall structure of the developed deep learning model followed by explaining the two improved components including aggregated residual block and locality sensitive hashing attention, as well as the training strategy.
The overall flowchart is illustrated in Fig. \ref{fig:model}.

\subsection{Overview}
\label{sec:unet}
U-Net was first proposed by Ronneberger et al.~\cite{ronneberger2015u}, which was basically a variant of fully convolutional networks (FCN)~\cite{long2015fully}. The traditional U-Net is a type of artificial neural network (ANN) containing a set of convolutional layers and deconvolutional layers to perform the task of biomedical image segmentation. The structure of U-Net is symmetric with two parts: encoder and decoder. The encoder is designed to extract the spatial features from the original medical image. The decoder is to construct the segmentation map from the extracted spatial features. The encoder follows the similar style like FCN with the combination of several convolutional  layers. To be specific, the encoder consists of a sequence of blocks for down-sampling operations, with each block including two $3\times 3$ convolution layers followed by a $2 \times 2$ max-pooling layers with stride of 2. The number of filters in the convolutional layers is doubled after each down-sampling operation. In the end, the encoder adopts two $3\times 3$ convolutional layers as the bridge to connect with the decoder.

Differently, the decoder is designed for up-sampling and constructing the segmentation image. The decoder first utilizes the a $2\times 2$ deconvolutional layer to up-sample the feature map generated by the encoder. The deconvolutional layer developed by Zeiler et al.~\cite{zeiler2010deconvolutional} contains the transposed convolution operation and will half the number of filters in the output. It is followed by a sequence of up-sampling blocks which consist two $3\times 3$ convolution layers and a deconvolutional layer. Then, a $1\times 1$ convolutional layer is used as the final layer to generate the segmentation result. The final layer adopted Sigmoid function as the activation function while all other layers used ReLU function.The ReLU and the Sigmoid functions are defined as:
\begin{align*}
    & \text{ReLU: } f(x)= \max\{0,x\}
  \\
  & \text{Sigmoid: } f(x) = \frac{1}{1+\exp(-x)}
\end{align*}
In addition, the U-Net concatenates part of the encoder features with the decoder. For each block in encoder, the result of the convolution before the max-pooling is transferred to decoder symmetrically. In decoder, each block receives the feature representation learned from encoder, and concatenates them with the output of deconvolutional layer. The concatenated result is then forwardly propagated to the consecutive block. This concatenation operation is useful for the decoder to capture the possible lost features by the max-pooling.

\subsection{Aggregated Residual Block}
\label{sec:encode}
\begin{figure}[h]
    \centering
    \includegraphics[width=0.7\linewidth]{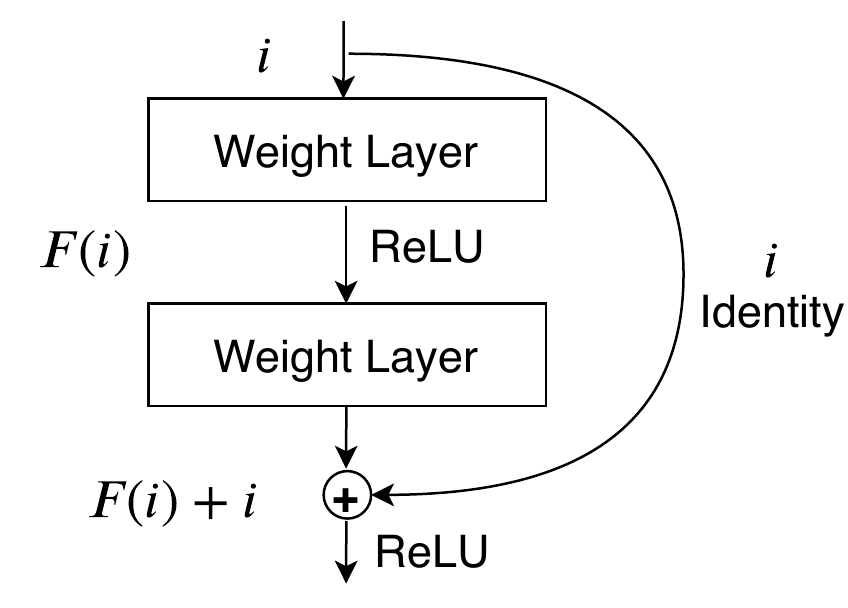}
    \caption{ResNet Block. Here the variable $i$ is the D-dimension representation of the input image or features map. The skip connection is performed as the identity mapping, the output of the identity mapping will be added to the output of the stacked layers.}
    \label{fig:res}
\end{figure}

As mentioned in previous section, the U-Net only have four blocks of convolution layers to conduct the feature extraction. The conventional structure may not be sufficient for the complex medical image analysis such as multi-class image segmentation in lung, which is the aim for this study. Although U-Net can easily separate the lung in a CT image, it may have limited ability to distinguish the difference infection regions of the lung which infected by COVID-19. Based on this case, the deeper network is needed with more layers, especially for the encoding process. However, when deeper network converging, a problem will be exposed: with increasing of the network depth, accuracy gets very high and then decrease rapidly. This problem is be defined as degradation problem~\cite{he2015convolutional,srivastava2015highway}. He et al. proposed the ResNet~\cite{he2016deep} to mitigate the effect of network degradation on model learning. ResNet utilizes a skip connection with residual learning to overcome the degradation and avoid estimating a large number parameters generated by the convolutional layer. The typical ResNet block is depicted as Fig. \ref{fig:res}.
The function $F$ can be defined as:
\begin{align*}
    F(i) = \sum_{j=1}^D w_ji_j
\end{align*}
where $i = [i_1, i_2, \cdots, i_D]$ and $W= [w_1, w_2, \cdots, w_D]$ is the trainable weight for the weight layer. Different from the U-Net that concatenates the features map into decoding process, ResNet adopts the shortcut to add the identity into the output of each block. The stacked residual block is able to better learn the latent representation of the input CT image. However, the model comes more complex and hard to converge as the increase in the number of layers. 

\begin{figure}[h]
    \centering
    \includegraphics[width=\linewidth]{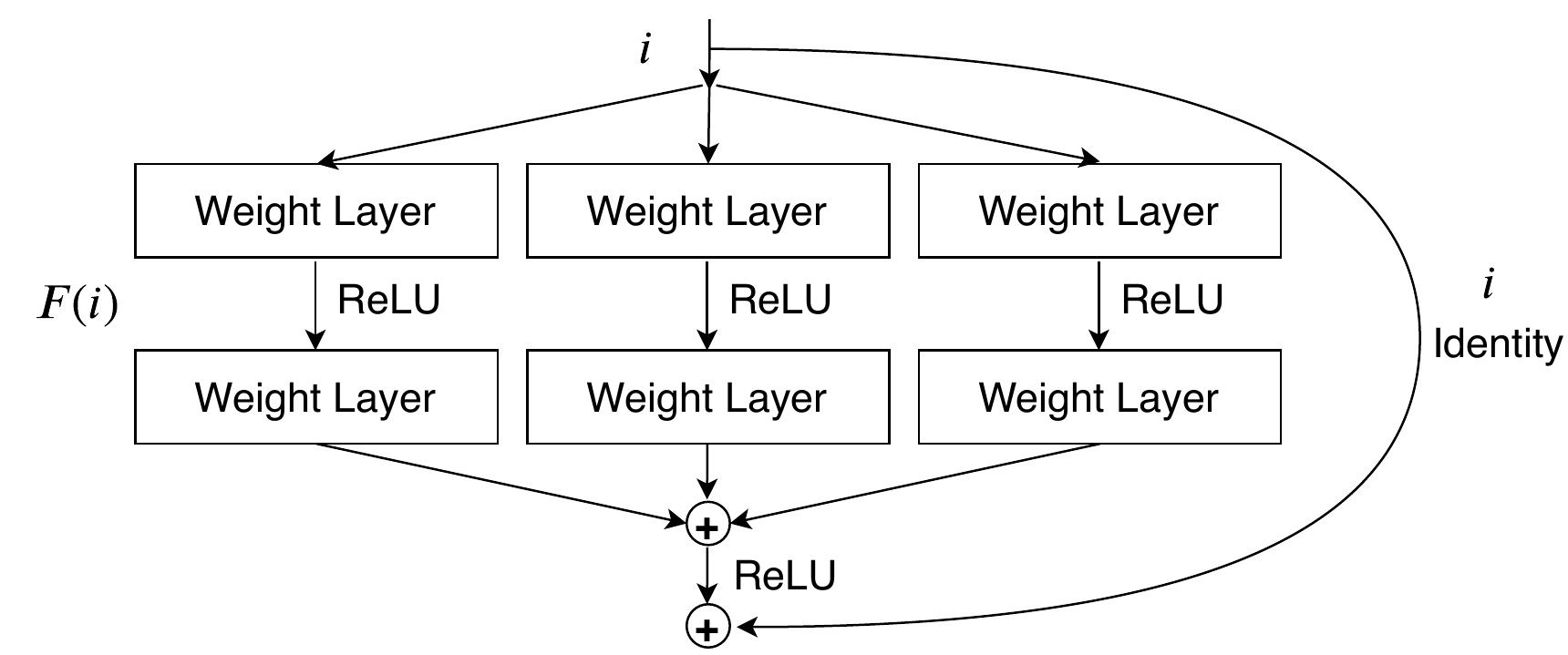}
    \caption{ResNeXt Block. The variable $i$ is the D-dimension representation of the input image or features map. Here, the cardinality = 3.}
    \label{fig:resnext}
\end{figure}

Regarding this, Xie et al. proposed Aggregated Residual Network(ResNeXt) and showed that increasing the cardinality was more useful than increasing the depth or width~\cite{xie2017aggregated}. The cardinality is defined as the set of the Aggregated Residual transformations with formulation as follows:
\begin{align*}
    F(i) = \sum_{j=1}^C \mathcal{T}_j(i)
\end{align*}
where $C$ is the number of residual transformation to be aggregated and $\mathcal{T}_j(i)$ can be any function. Considering a simple neuron, $\mathcal{T}_j$ should be a transformation projecting $i$ into an low-dimensional embedding ideally and then transforming it. Accordingly, we can extend it into the residual function:
\begin{align*}
    y = \sum_{j=1}^C \mathcal{T}_j(i) + i
\end{align*}
where the $y$ is the output. The ResNeXt block is visualized in Fig. \ref{fig:resnext}. Compared with the Fig. \ref{fig:res}, the ResNeXt has a slightly different structure. The weight layer's size is smaller than ResNet as ResNeXt use the cardinality to reduce the number of layers but keep the performance. One thing is wroth to mention that the three small blocks inside the ResNeXt block need to have the some topology, in the other words, they should be topologically equivalent. 

Similar with the ResNet, after a sequence of blocks, the learned features are feed into a global averaging pooling layers to generate the final feature map. Different from the convolutional layers and normal pooling layers, the global averaging pooling layers take the average of feature maps derived by all blocks. It can sum up all the spatial information which captured by each step and is generally more robust than directly make the spatial transformation to the input. Mathematically, we can treat the global averaging pooling layer as a structural regularizer that are helpful for driving the desired feature maps~\cite{zhou2016learning}.

Importantly, instead of using the encoder in the U-Net, our proposed deep learning model adopts the ResNeXt block (see Fig. \ref{fig:resnext}) to conduct the features extraction. The ResNeXt provides a solution which can prevent the network goes very deeper but remain the performance. In addition, the training cost of ResNeXt is better than ResNet.

\subsection{Locality Sensitive Hashing Attention}
\label{sec:attention}
\begin{figure}
    \centering
    \includegraphics[width=0.9\linewidth]{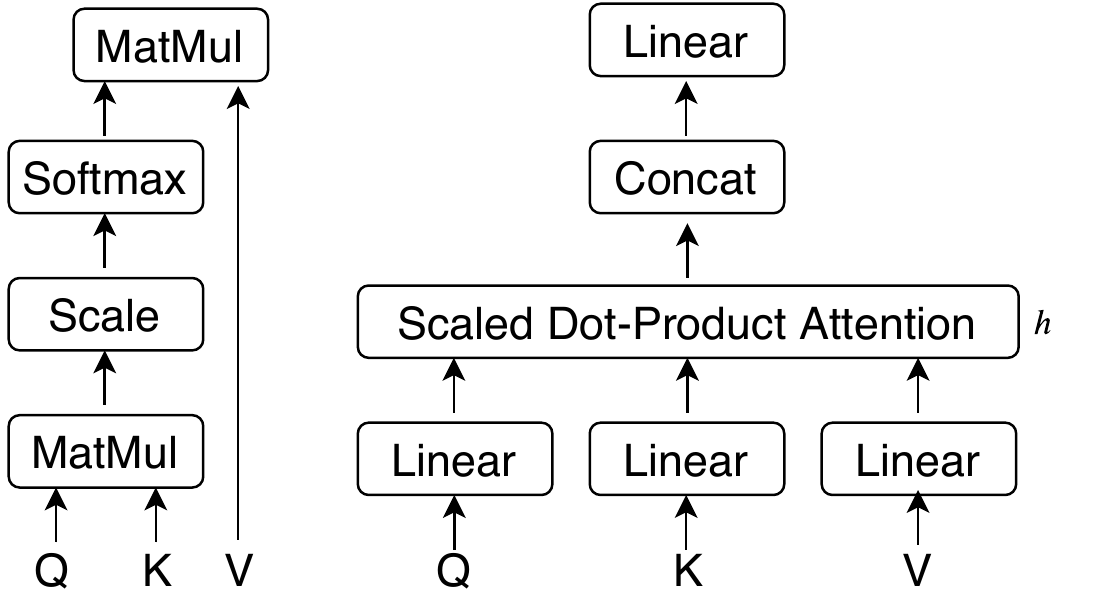}
    \caption{Attention Mechanism. The left figure shows the simple scaled dot-prodct attention. The right figure depicts the multi-hand attention with the $h$ head.}
    \label{fig:att}
\end{figure}
The decoder in U-Net is used to up-sampling the extracted feature map to generate the segmentation image. However, due to the capability of the convolutional neural network, it may not able to capture the complex features if the network structure is not deep enough. In recent years, transformers~\cite{sutskever2014sequence} have gained increasingly interest~\cite{devlin2018bert}. The key of the success is the attention mechanism~\cite{vaswani2017attention}. Attention includes two different mechanisms: soft attention and hard attention. We adopt the soft attention to improve the model learning. Different the hard attention, the soft attention can let model focus on each pixel's relative position, but the hard attention only can focus on the absolute position. There are two different types of soft attention: Scaled Dot-Product Attention and Multi-Head Attention as shown in Fig. \ref{fig:att}. The scaled dot-product attention takes the inputs including a query $Q$, a key $K_n$ of the $n$-dimension and a value $V_m$ of the $m$-dimension. 
The dot-product attention is defined as follows:
\begin{align*}
    \text{Attention}(Q,K_n,V_m) = \text{softmax}(\frac{QK_n^T}{\sqrt{n}})V_m
\end{align*}
where $K_n^T$ represent to the transpose of the matrix $K_n$ and $\sqrt{n}$ is a scaling factor. The softmax function $\sigma(\textbf{z})$ with $\textbf{z} = [z_1, \cdots, z_n]\in \mathbb{R}^n$ is given by:
\begin{align*}
    \sigma(\textbf{z})_i = \frac{\exp(z_i)}{\sum_{j=1}^n \exp(z_j)} \text{ for }i = 1,\cdots, n 
\end{align*}
Vaswani et al.~\cite{vaswani2017attention} mentioned that, perform different linearly project of the queries $Q$, keys $K$ and values $V$ in parallel $h$ layers will benefit the attention score calculation. We can assume that $Q,K$ and $V$ have been linearly projected to $d_k, d_k, d_v$ dimensions, respectively. It is worth noting that these linear projections are different and learnable. On each projection $p$, we have a pair of query, key and value $Q_p,K_p, V_p$ to conduct the attention calculation in parallel, which results in a $d_v$-dimensional output. The calculation can be formulated as:
\begin{align*}
    \text{MultiHead}(Q,K,V) = \text{Concatenate}(\text{head}_1, \cdots, \text{head}_h)W^O \\
    \text{where head}_i = \text{Attention}(QW_i^Q,KW_i^K,VW_i^V) 
\end{align*}
where the the projections $W_i^Q \in \mathbb{R}^{d_{model} \times d_k}$, $W_i^K \in \mathbb{R}^{d_{model} \times d_k}$, $W_i^V\in \mathbb{R}^{d_{model} \times d_v}$are parameter matrices  and $W^O \in \mathbb{R}^{d_{model} \times hd_v}$ is the weight matrix used to balance the results of $h$ layers.

However, the multi-head attention is memory inefficient due to the size of $Q,K$ and $V$. Assume that the $Q,K,V$ have the shape $[|batch|, length, d_{model}]$ where $|\cdot|$ represents the size of the variable. The term $QK^T$ will produce a tensor in shape $[length, length, d_{model}]$. Given the standard image size, the length $\times$ length will take most of the memory. Kitaev et al.~\cite{kitaev2020reformer} proposed a Locality Sensitive Hashing(LSH) based Attention to address this issue. Firstly, we rewire the basic attention formula into each query position $i$ in the partition form:
\begin{align*}
    a_i = \sum_{j\in P_i}\frac{\exp(q_i \cdot k_j - z(i,P_i))v_j}{\sqrt{d_k}} \text{ where } P_i = \{j:i\geq j\}
\end{align*}
where the function $z$ is the partition function, $P_i$ is the set which query position $i$ attends to. During model training, we normally conduct the batching and assume that there is a larger set $P_i^L = \{0,1,\cdots,l\}\supseteq P_i$ without considering elements not in $P_i$:
\begin{align}
    a_i = \sum_{j\in P_i^L}\frac{\exp(q_i \cdot k_j - N(j,P_i) - z(i,P_i))v_j}{\sqrt{d_k}} \\ 
    \text{ where } N(j,P_i) = 
    \begin{cases} 
      0 & j \in P_i \\
      \infty & j \notin P_i
   \end{cases} 
\end{align}
Then, with a hash function $h(\cdot)$: $h(q_i) = h(k_j)$, we can get $P_i$ as:
\begin{align*}
    P_i = \{j:h(q_i) = h(k_j)\}
\end{align*}
In order to guarantee that the number of keys can uniqually match with the number of quires, we need to ensure that $h(q_i) = h(k_i)$ where $k_i = \frac{q_i}{\|q_i\|}$. During the hashing process, some similar items may fall in different buckets because of the hashing. The multi-round hashing provides an effective way to overcome this issue. Suppose there is $n_r$ round, and each round has different hash functions $\{h_1, \cdots, h_{n_r}\}$, so we have:
\begin{align}
    P_i = \bigcup_{g=1}^{n_r} P_i^{g} \text{ where } P_i^g = \{j:h^g(q_i) = h^g(q_j)\}
\end{align}
Considering the batching case, we need to get the $P_i^L$ for each round $g$:
\begin{align*}
    \widehat{P_i^L} = \{j:\lfloor\frac{i}{m}\rfloor -1 \leq \lfloor\frac{j}{m}\rfloor \leq \lfloor\frac{i}{m}\rfloor\}
\end{align*}
where $m=\frac{2l}{n_r}$. The last step is to calculate the LSH attention score in parallel. With the formula (1) and (3), we can derive:
\begin{align*}
    & a_i = \sum_{g=1}^{n_r} \frac{\exp(z(i,P_i^g)-z(i,P_i))a_i^g}{\sqrt{d_k}} \\
    & \text{where } a_i^g = \sum_{j\in \widehat{P_i^L}} \frac{\exp(q_i\cdot k_j - m_{i,j}^g -z(i,P_i^g))v_j}{\sqrt{d_k}}    \\
    & \text{with }m_{i,j}^g = 
    \begin{cases} 
      \infty & j \notin P_i^g \\
      10^5 & i = j \\
      \log |\{g': j \in P_i^{g'}\}| & otherwise
   \end{cases} 
\end{align*}

\subsection{Training Strategy}
The task of the lung CT image segmentation is to predict if each pixel of the given image belongs to a predefined class or the background. Therefore, the traditional medical image segmentation problem comes to a binary pixel-wise classification problem. However, in this study, we are focusing on the multi-class image segmentation, which can be concluded as a multi-classes pixel-wise classification. Hence, we choose the multi-class cross entropy as the loss function:
\begin{align*}
  \mathcal{L} =  - \sum_{c=1}^M y_{o,c}\log(p_{o,c})
\end{align*}
where $y_{o,c}$ is a binary value which use to compare the correct class $c$ and observation class $o$, $p_{o,c}$ is a probability of the observation $o$ to correct class $c$ and $M$ is the number of classes. 

\section{Experiment and Evolution Result}
\subsection{Data Description}
\begin{figure*}[ht]
    \begin{subfigure}{0.245\linewidth}
        \includegraphics[width=\linewidth]{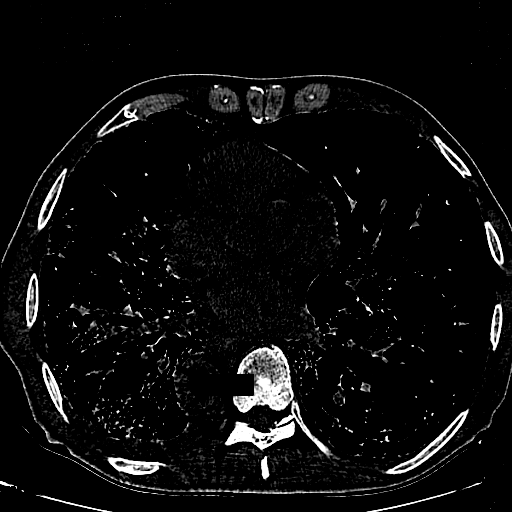}
        \caption{}
    \end{subfigure}
    \begin{subfigure}{0.245\linewidth}
        \includegraphics[width=\linewidth]{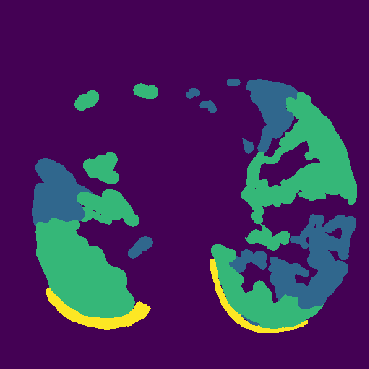}
        \caption{}
    \end{subfigure}
    \begin{subfigure}{0.245\linewidth}
        \includegraphics[width=\linewidth]{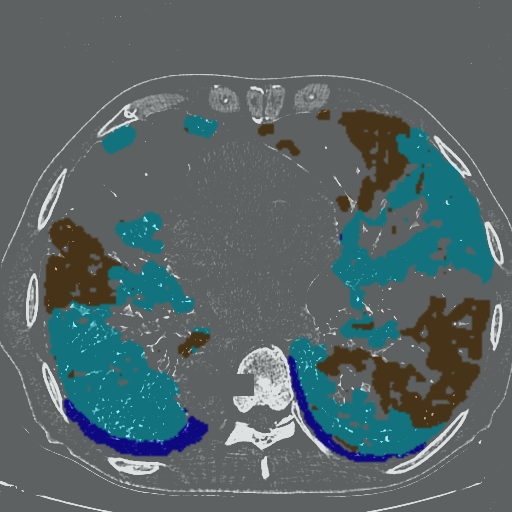}
        \caption{}
    \end{subfigure}
    \begin{subfigure}{0.245\linewidth}
        \includegraphics[width=\linewidth]{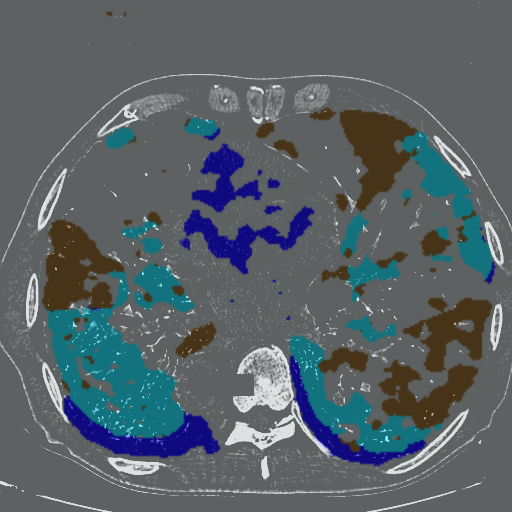}
        \caption{}
    \end{subfigure}
    \begin{subfigure}{0.245\linewidth}
        \includegraphics[width=\linewidth]{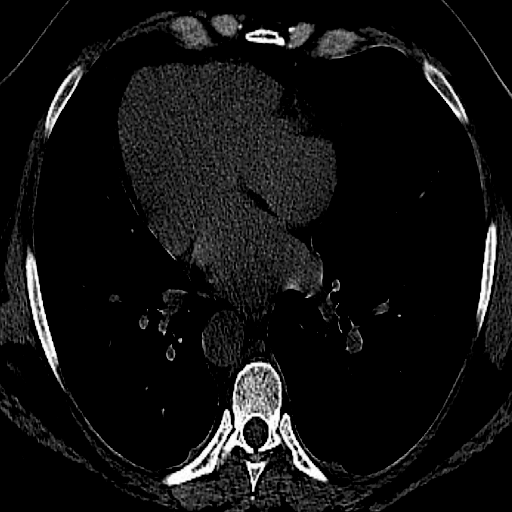}
        \caption{}
    \end{subfigure}
    \begin{subfigure}{0.245\linewidth}
        \includegraphics[width=\linewidth]{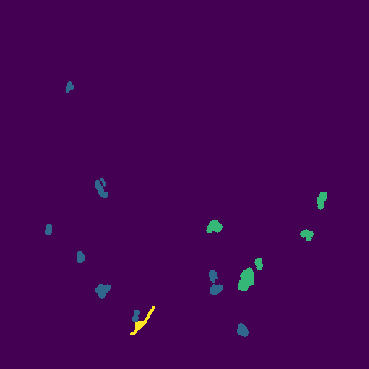}
        \caption{}
    \end{subfigure}
        \begin{subfigure}{0.245\linewidth}
        \includegraphics[width=\linewidth]{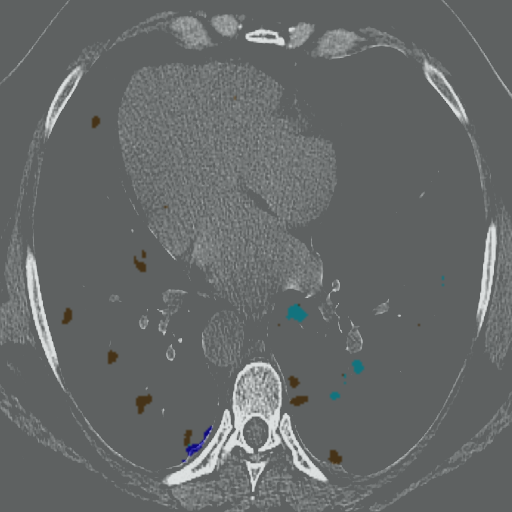}
        \caption{}
    \end{subfigure}
    \begin{subfigure}{0.245\linewidth}
        \includegraphics[width=\linewidth]{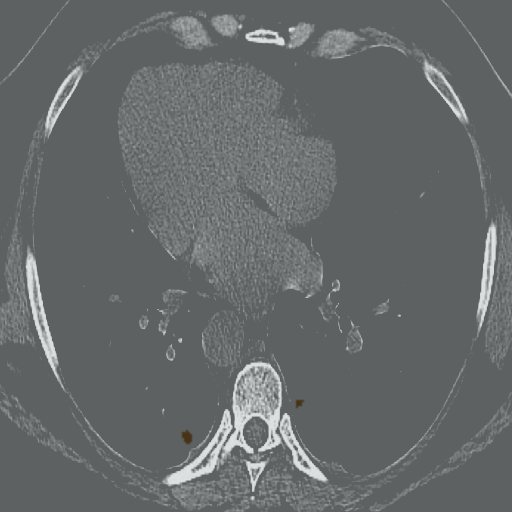}
        \caption{}
    \end{subfigure}
    \caption{Visualization of segmentation results. The images (a) and (e) show the preprocessed chest CT images of two scans. The images (b) and (f) are the ground-truth masks for these two scans, where the yellow represents the consolidation, blue represents pleural effusion and green corresponds to ground-glass opacities. The images (c) and (g) are the segmentation results generated by our model where the blue represents the consolidation and brown represents the pleural effusion and sky-blue for the ground-glass opacities. The images (d) and (h) are the outputs of the U-Net. In order to make the visualization clear, we choose the light gray as the color for the background segment.}
    \label{fig:result}
\end{figure*}
We used COVID-19 CT images collected by Italian Society of Medical and Interventional Radiology (SIRM)\footnote{https://www.sirm.org/category/senza-categoria/covid-19/} for our experimental study. The dataset included 110 axial CT images collected from 60 patients. These images were reversely intensity-normalized by taking RGB-values from the JPG-images from areas of air (either externally from the patient or in the trachea) and fat (subcutaneous fat from the chest wall or pericardial fat) and used to establish the unified Houndsfield Unit-scale (the air was normalized to -1000, fat to -100). The ground-truth segmentation was done by a trained radiologist using MedSeg\footnote{http://medicalsegmentation.com/} with three labels: 1 = ground class opacification, 2 = consolidations, and 3 = pleural effusions. A total of 100 samples that have both preprocessed CT images and masks were used for our experimental analysis. These data are publicly available\footnote{http://medicalsegmentation.com/covid19/}.

\subsection{Data Preprocessing and Augmentation}
The original CT images have the size of 512 $\times$ 512. We use the opencv2\footnote{https://opencv.org/opencv-2-4-8/} to convert the images into size of 369 $\times$ 369 and grey scale. This processing is helpful to automatically minimize the effects of the black frame in the images and some random noises (e.g., words) on the segmentation.

As our model is based on deep learning, the number of samples will affect the performance significantly. Consider about the size of the dataset, data augmentation is necessary for training the neural network to achieve high generalizability. Our study implements parameterized transformations to realize data augmentation in this study. We rotate the existing images 90 degrees, 180 degrees and 270 degrees to generate another 300 examples. We can easily generate the corresponding mask by rotating with the same degrees. Scaling have the some property with the rotation, so we just scale the image to 0.5 and 1.5 separately to generate another 200 images and its corresponding masks.

\subsection{Experiments setting and Measure Metrics}
\label{sec:exp}
For the model training, we use the Adma~\cite{kingma2014adam} as the optimizer. 
For a fair comparison, we train our model and the U-Net with the default parameter in 100 epochs. Both models are trained under data augmentation and non-augmentation cases. We conducted the experimental analyses on our own server consisting of two 12-core/ 24-thread Intel(R) Xeon(R) CPU E5-2697 v2 CPUs, 6 NVIDIA TITAN X Pascal GPUs, 2 NVIDIA TITAN RTX, a total 768 GiB memory. 

In a segmentation task, especially for the multi-class image segmentation, the target area of interest may take a trivial part of the whole image. Thus, we adopt the Dice Score, accuracy, and precision as the measure metrics. The dice score is defined as:
\begin{align*}
    DSC(X,Y) = \frac{2|X\cap Y|}{|X|+|Y|}
\end{align*}
where $X$, $Y$ are two sets, and $|\cdot|$ calculates the number of element in a set. Assume $Y$ is the correct result of the test and $X$ is the predicted result.
We conduct the experimental comparison based on a 10-fold cross-validation for performance assessment. 

\subsection{Results}
The figure \ref{fig:result} provides two examples about the result images which have data augmentation. The table \ref{tab:re1} shows the measure metric for our proposed model and the U-Net in with data augmentation case and no data augmentation case.
\begin{table}[ht]
    \centering
     \caption{Comparison of segmentation performance between our proposed model and U-Net. All the values are the average value based on the 10-fold cross-validation.}
    \begin{tabular}{c|c|c|c|c|c|c}
         Model & \multicolumn{3}{c|}{With Augmentation}& \multicolumn{3}{c}{No Augmentation}  \\ \hline
         & DSC & Acc & Precision & DSC & Acc & Precision  \\ \hline
         Ours & 0.94 & 0.89 & 0.95 & 0.83 & 0.79 & 0.82  \\ \hline
         U-Net & 0.82 & 0.79 & 0.83 & 0.75 & 0.70 & 0.72 \\ \hline
         Improve & 14.6\% & 12.7\% & 14.5\% & 10.7\% & 12.9\% & 13.9\% \\\hline
    \end{tabular}
    \label{tab:re1}
\end{table}
Based on this table, we can easily find that our proposed method is out-performed than U-Net which the improvement is at least 10\% in all three measure metrics. 
As shown in figure \ref{fig:res}(h), we find that the original U-Net almost failed to do the segmentation. The most possible reason is that, the range of interest is very small, and the U-Net do not have enough capability to distinguish those trivial difference.

\subsection{Ablation Study}
In addition to the above-mentioned results, we are also interested in the effectiveness of each component in the proposed model. Accordingly, we conduct the ablation study about the ResNeXt and Attention separately to investigate how these components would affect the segmentation performance. To ensure a fair experimental comparison, we conduct the ablation study in the exactly same experiment environment with our main experiments presented in section \ref{sec:exp}. We implement the ablation study on two variants of our model: Model without Attention and Model without ResNeXt. Our model without ResNeXt is similar with literature~\cite{oktay2018attention}. We just use the M-R to represent it. The results are summarized in Table \ref{tab:re2}, where M-A represents the model without attention and M-R represents the model without ResNeXt block. We can observe that both the attention and ResNeXt blocks play important roles in our model and contribute to derive improved segmentation performance in comparison with U-Net.
\begin{table}[h]
    \centering
     \caption{Comparison result of ablation study. All the values are the average value based on the 10-fold cross-validation.}
    \begin{tabular}{c|c|c|c|c|c|c}
         Model & \multicolumn{3}{c|}{With Augmentation}& \multicolumn{3}{c}{No Augmentation}  \\ \hline
         & DSC & Acc & Precision & DSC & Acc & Precision  \\ \hline
         Ours(M) & 0.94 & 0.89 & 0.95 & 0.83 & 0.79 & 0.82  \\ \hline
         M - A & 0.85 & 0.82 & 0.84 & 0.79 & 0.74 & 0.77 \\ \hline
         M - R & 0.84 & 0.81 & 0.83 & 0.77 & 0.76 & 0.77 \\\hline
         U-Net & 0.82 & 0.79 & 0.83 & 0.75 & 0.70 & 0.72 \\ \hline
    \end{tabular}
    \label{tab:re2}
\end{table}

\section{Discussion and Conclusions}
Up to now, the most common screening tool for COVID-19 is the CT imaging. It can help community to accelerate the speed of diagnose and accurately evaluate the severity of COVID-19~\cite{shan+2020lung}. In this paper, we presented a novel deep learning-based algorithm for automated segmentation of COVID-19 CT images, which is proved to be plausible and superior comparing to a series of baselines. We proposed a modified U-Net model by exploiting residual network to enhance the feature extraction. An efficient attention mechanism was further embedded into the decoding process to generate the high-quality multi-class segmentation results. Our method gained more than 10\% improvement in multi-class segmentation when comparing against U-Net and a set of baselines.

Recent study shows that the early detection of the COVID-19 is very important~\cite{bernheim2020chest}. If the infection in chest CT image can be detected at early stage, the patients would have the higher chance to survive~\cite{song2020emerging}. Our study provides an effective tool for the radiologist to precisely determine the lung's infection percentage and diagnose the progression of COVID-19. It also shed some light on how deep learning can revolutionize the diagnosis and treatment in the midst of COVID-19.  

Our future work would be generalizing the proposed model into a wider range of practical scenarios, such as facilitating with diagnosing more types of diseases from CT images. In particularly, in the case of a new disease, such as the coronavirus, the amount of ground truth data is usually limited given the difficulty of data acquisition and annotation. The model is capable of generalizing and adapting itself usingonly a few available ground-truth samples. A knowledge-based generative model \cite {zhang2019adversarial} will be integrated to enhance the ability in handling new tasks. Another line of future work lies in the interpretability, which is specially critical for the medical domain applications. Although deep learning is widely accepted to its limitation in interpretability, the attention mechanism we proposed in this work can produce the interpretation of internal decision process at some levels. To gain deeper scientific insights, we will keep working along with this direction and explore the hybrid attention model for generating meaningfully semantic explanations. 

\bibliographystyle{IEEEtran}
\bibliography{IEEEabrv,sample}
\end{document}